\documentclass[%
reprint,
superscriptaddress,
nofootinbib,
 amsmath,amssymb,footinbib,
 aps,
 prb,
]{revtex4-2}

\usepackage[utf8]{inputenc}
\usepackage{graphicx}
\usepackage{placeins}
\usepackage{float}
\usepackage{subfigure}
\usepackage{longtable}
\usepackage{amssymb}
\usepackage{natbib}
\usepackage{xcolor}
\usepackage{hyperref}
\usepackage{amsmath}
\usepackage{siunitx}
\usepackage{xcolor}
\usepackage{csquotes}
\usepackage{nicefrac}
\usepackage[cal=boondoxo]{mathalpha}

\newcommand\enn{\mathcal{n}}
\newcommand\emm{\mathcal{m}}

\begin{document}


\title{Engineering {\color{black} underdoped} CuO$_2$ nanoribbons in nm-thick $a$-axis YBa$_2$Cu$_3$O$_{7-\delta}$ films}

\author{Riccardo Arpaia}
 \email{riccardo.arpaia@chalmers.se}
\affiliation{Quantum Device Physics Laboratory, Department of Microtechnology and Nanoscience, Chalmers University of Technology, SE-41296, Göteborg, Sweden}
\author{Núria Alcalde-Herraiz}
\affiliation{Quantum Device Physics Laboratory, Department of Microtechnology and Nanoscience, Chalmers University of Technology, SE-41296, Göteborg, Sweden}
\author{Andrea D'Alessio}
\altaffiliation[Present address: ]{Dept of Energy Conversion and Storage, Technical University of Denmark, DK-2800, Kgs. Lyngby, Denmark}
\affiliation{Quantum Device Physics Laboratory, Department of Microtechnology and Nanoscience, Chalmers University of Technology, SE-41296, Göteborg, Sweden}
\affiliation{Dipartimento di Fisica, Politecnico di Milano, piazza Leonardo da Vinci 32, I-20133 Milano, Italy}
\author{Evgeny Stepantsov}
\affiliation{Quantum Device Physics Laboratory, Department of Microtechnology and Nanoscience, Chalmers University of Technology, SE-41296, Göteborg, Sweden}
\affiliation{Shubnikov Institute of Crystallography of the Federal Scientific Research Centre ``Crystallography and Photonics" of the Russian Academy of Sciences, Leninskiy pr. 59, Moscow RU-119333, Russia}
\author{Eric Wahlberg}
\affiliation{Quantum Device Physics Laboratory, Department of Microtechnology and Nanoscience, Chalmers University of Technology, SE-41296, Göteborg, Sweden}
\affiliation{RISE Research Institutes of Sweden, Box 857, SE-50115, Borås, Sweden}
\author{Alexei Kalaboukhov}
\affiliation{Quantum Device Physics Laboratory, Department of Microtechnology and Nanoscience, Chalmers University of Technology, SE-41296, Göteborg, Sweden}
\author{Thilo Bauch}
\affiliation{Quantum Device Physics Laboratory, Department of Microtechnology and Nanoscience, Chalmers University of Technology, SE-41296, Göteborg, Sweden}
\author{Floriana Lombardi}%
 \email{floriana.lombardi@chalmers.se}
\affiliation{Quantum Device Physics Laboratory, Department of Microtechnology and Nanoscience, Chalmers University of Technology, SE-41296, Göteborg, Sweden}%

\date{\today}
\begin{abstract}
In underdoped cuprate high $T_{\mathrm{c}}$ superconductors, various local orders and symmetry breaking states, in addition to superconductivity, reside in the CuO$_2$ planes. The  confinement  of the CuO$_2$ planes can therefore  play a fundamental role in modifying the hierarchy between the various orders and their intertwining with superconductivity.  Here we present the growth of $a$-axis oriented YBa$_2$Cu$_3$O$_{7-\delta}$ films, spanning the whole underdoped side of the phase diagram. In these samples, the CuO$_2$ planes are confined by the film thickness, effectively forming unit-cell-thick nanoribbons. The unidirectional confinement at the nanoscale enhances the in-plane anisotropy of the films. By X-ray diffraction and resistance vs temperature measurements, we have discovered the suppression of the orthorhombic-to-tetragonal transition at low dopings, and a very high anisotropy of the normal state resistance in the $b$-$c$ plane, the latter being connected to a weak coupling between adjacent CuO$_2$ nanoribbons.  These findings show that the samples we have grown represent a novel system, different from the bulk, where future experiments can possibly shed light on the rich and mysterious physics occurring within the CuO$_2$ planes.
\end{abstract}

\maketitle

\section{Introduction}
\begin{figure}[b]
\centering
\includegraphics[width=8.5cm]{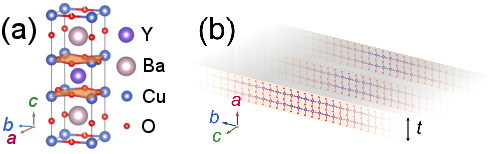}
\caption{\label{fig:Fig1} CuO$_2$ nanoribbons from YBCO. (a) YBCO unit cell. The CuO$_2$ planes where superconductivity, as well as various electronic local orders, reside, are highlighted in orange. (b) By growing $a$-axis oriented YBCO films, the   CuO$_2$ planes form ribbons of nanometer width, confined in one direction by the film thickness $t$. 
} 
\end{figure}

Cuprate high critical temperature superconductors (HTS) represent a prominent example of compounds dominated by strong electron-electron correlations. Here all the relevant physics takes place in planes, where a single copper atom is at the centre of a square of oxygen ligands. Despite this apparent simplicity, these materials host some of the most intricate quantum phases, dominated by strong interactions between spin, charge and orbital degrees of freedom, which are controlled by the element atomic positions within the crystal lattice \cite{keimer2015quantum}. Unraveling such intricate ground state is a grand challenge: the large electron-electron repulsion and the consequent short-range antiferromagnetic correlations, within the low dimensional CuO$_2$ planes (see orange surfaces in Fig. \ref{fig:Fig1}(a)), 
render  HTS systems intrinsically different from standard metals, generally well understood in terms of the Landau Fermi liquid paradigm. 
The resulting phase diagram, as a function of the doping and of the temperature, is very rich and complex,
hosting many intertwined and/or competing local ordering phenomena breaking the rotational/translational symmetry \cite{fradkin2015colloquium, loret2019intimate}. Examples of these orders, all residing within the CuO$_2$ planes, are charge density waves (CDW) \cite{ghiringhelli2012long, chang2012direct, gerber2015three} and charge density fluctuations (CDF) \cite{arpaia2019dynamical, yu2020unusual, arpaia2021charge, huang2021quantum, arpaia2023signature}, spin density waves (SDW) \cite{haug2010neutron, li2011magnetic}, electronic nematicity \cite{kivelson1998electronic, daou2010broken, wu2017spontaneous}, orbital currents \cite{fauque2006magnetic, varma2020colloquium} and possibly an unconventional spatially modulated Cooper pair density \cite{hamidian2016detection, agterberg2020physics}. 

Because of this complexity, new experimental strategies are needed to control the interactions and emergent properties and to establish the  hierarchy between cause and effect among the different orders \cite{ahn2021designing}.

A common approach, used to tune the local properties of cuprate HTS, is to grow them in thin film form. The epitaxial strain induced by the substrate has been used to control the atomic positions and therefore the strengths of the interactions.  More in general, strain allows to control the critical temperature \cite{locquet1998doubling}, as well as the magnetic-exchange interaction \cite{ivashko2019strain}, the nematicity \cite{wu2017spontaneous, nakata2021nematicity} and the charge density waves \cite{kim2018uniaxial, bluschke2018stabilization, wahlberg2021restored, choi2022unveiling, gupta2023tuning}. 

The ultimate manipulation of the HTS cuprates could be represented by the confinement of the  CuO$_2$ planes on the nanoscale - a realm which is still largely unexplored.  In this limit, possibly degenerate ground states, different from those of the corresponding bulk, could become favorable. 
A way to confine the  CuO$_2$ planes is to grow high quality $a$-axis oriented thin films. In $a$-axis films, the CuO$_2$ planes, hosting superconductivity, are constrained in one direction by the thickness of the film and are hence shaped into a quasi-one-dimensional geometry, effectively forming CuO$_2$ ribbons of nanometer width, perpendicular to the substrate (see Fig. \ref{fig:Fig1}(b)). One can expect that this confinement influences the properties of the nanoscale ordering phenomena, whose characteristic lengths are comparable with the lateral dimensions of the CuO$_2$ ribbons \cite{tranquada1995evidence, ghiringhelli2012long, hamidian2016detection}.  

The growth of high-quality epitaxial $a$-axis films is however challenging: differently from the $c$-axis counterpart, $a$-axis grains nucleate at rather low deposition temperatures \cite{miletto2000competition}, which favor the growth of less homogeneous films, with poor crystallinity \cite{hwang1990application}.  Despite considerable efforts to produce optimally-doped films and heterostructures of high quality \cite{eom1990epitaxial, luo1991axis, inam1991properties, hontsu1992axis, mukaida1992growth, hamet1992axis, mukaida1993plane, suzuki1993anisotropic, ito1994superconductivity, mahajan1994growth, park1994growth, sung1995superconducting, trajanovic1996grain, trajanovic1997resistivity, fuchs1997growth, baghdadi2017study, sueyoshi2019flux, suyolcu2021axis}, the impact of CuO$_2$ confinement on the cuprate phase diagram requires the growth of underdoped films, where the majority of local orders manifest \cite{keimer2015quantum, fradkin2015colloquium}. Therefore, achieving controlled variation of the doping level is crucial, while maintaining the anisotropic characteristics of the bulk material, i.e., growing fully detwinned samples. These tasks are challenging even for $c$-axis oriented films and have never been successfully executed with $a$-axis oriented films.

We have previously reported the growth of high quality optimally doped $a$-axis oriented YBa$_2$Cu$_3$O$_{7-\delta}$ (YBCO) thin films \cite{baghdadi2017study}.
In this work we present the properties of $a$-axis oriented YBCO thin films, at various doping levels from the optimally doped  down to the insulating state, obtained by a careful oxygen annealing procedure. X-ray diffraction (XRD) analysis shows that the films are fully detwinned, implying the presence of CuO chains, aligned throughout the sample along the in-plane $b$ direction. By varying the film thickness up to 800 nm, we have verified the dominant role of strain in this peculiar system. We find that the unit cell differs from the bulk even at dimensions of several hundreds of nanometers. The resistance vs temperature measurements are characterized by sharp superconducting transitions, showing that the superconducting properties of the films are rather homogeneous. We observe a strong in-plane anisotropy of the resistance, comparable, also for rather thick films, to the best results achieved on detwinned single crystals. This opens up the possibility of studying potential modifications of the ground state induced by confinement in films with only a few unit cells in thickness.


\section{Thin film growth and role of the oxygen annealing}

To favor the growth of $a$-axis oriented films, we have deposited by rf sputtering, at a power of 50 W and pressure of 0.1 mbar (Ar:O$_2$=4:1), a 90 nm thick film of PrBa$_2$Cu$_3$O$_{7-y}$ (PBCO), acting as an insulating buffer layer, on a (100) SrLaGaO$_4$ (SLGO) substrate \footnote {As previously shown in literature \cite{hontsu1992axis, sung1995superconducting, baghdadi2017study}, the usage of a PBCO buffer layer, on top of a (100) SLGO substrate, favors the growth of untwinned films, with the $a$-axis oriented along the out-of-plane direction. The length of the PBCO $b$-axis (3.91 \AA \, in bulk form) is indeed closer to the YBCO $b$-axis (3.89 \AA) than to the YBCO $a$-axis (3.82 \AA). It is therefore energetically more favorable for the YBCO $b$-axis to align along the PBCO $b$-axis, i.e. on the plane of the substrate/buffer layer.}. The temperature is gradually increased during the deposition from 660$^{\circ}$C, where the first nuclei are formed, up to 830$^{\circ}$C. After the deposition of this buffer layer, which is known to promote the growth of homogeneous $a$-axis superconducting films, with high crystallinity \cite{hontsu1992axis, mukaida1993plane}, the sample is transferred in-situ to a pulsed laser deposition chamber. Here, a YBCO film, with a thickness varying in the range 50-800 nm, is deposited at 805$^{\circ}$C and an oxygen pressure of 0.9 mbar.  By slowly decreasing the temperature after the deposition, at an oxygen pressure of 900 mbar, we achieve slightly overdoped films where the CuO chains are fully oxidized. 

\begin{figure}[h]
\centering
\includegraphics[width=8.5cm]{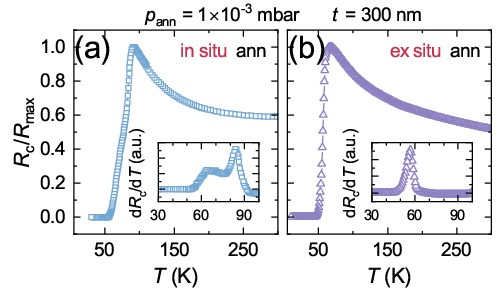}
\caption{\label{fig:Fig2} Dependence of the superconducting transition of $a-$axis YBCO thin films on the oxygen annealing. (a) The resistance measured along the $c$-axis direction, $R_c$, normalized to the maximum value $R_{\mathrm{max}}$, is plotted vs temperature for a 300 nm thick underdoped film, achieved using an in-situ annealing oxygen pressure $p_{\mathrm{ann}} = 1\times10^{-3}$ mbar. In the inset, the derivative of the $R_c$($T$) highlights the pronounced broadening of the superconducting transition. (b) $R_c$($T$) is presented for an underdoped film, obtained by applying  the same annealing pressure used for the film in panel (a), but ex-situ. The homogeneity of the film is proven by a rather sharp superconducting transition, as highlighted in the inset.
} 
\end{figure}

\begin{figure*}[bth]
\centering
\includegraphics[width=17.8cm]{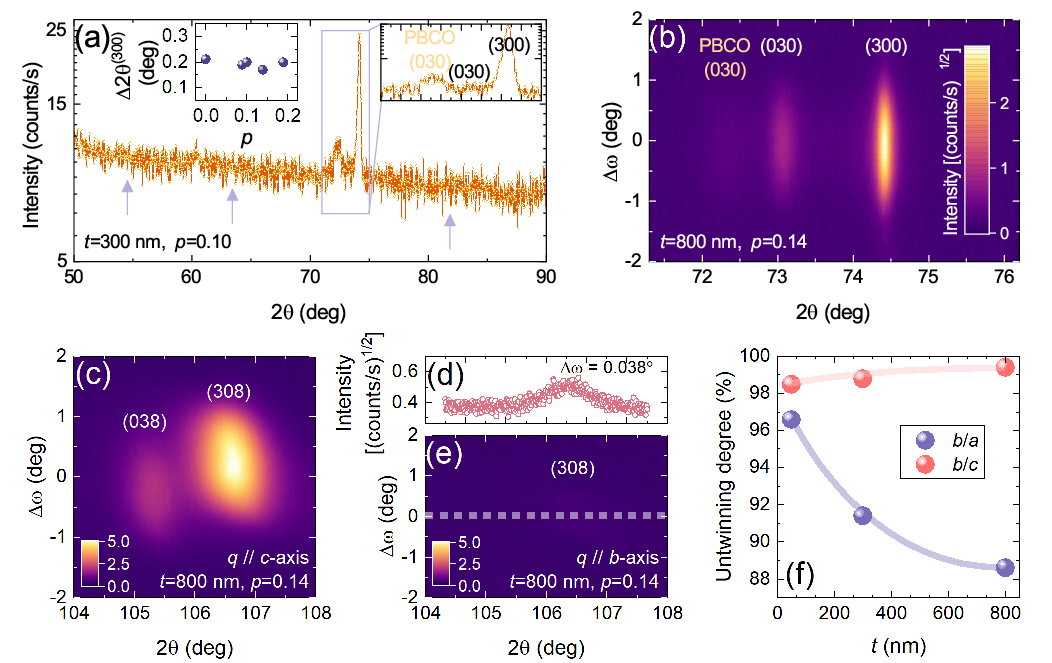}
\caption{\label{fig:Fig3} Complete detwinning in $a-$axis oriented YBCO films. (a) Symmetrical $2\theta-\omega$ scan of a 300 nm thick underdoped film. The arrows show the positions where the ($00\enn$) YBCO reflections should be for a $c$-axis oriented film. The inset on the left shows the FWHM of the (300) YBCO Bragg peak, $\Delta2\theta^{(300)}$, as a function of $p$. (b) Symmetrical $2\theta-\omega$ map of a 800 nm, slightly underdoped, YBCO film. Asymmetrical  $2\theta-\omega$ maps on the same sample are performed with the momentum $q$ parallel either to (c) the $c$-axis or to (d)-(e)the $b$-axis  direction. Panel d is the linescan, highlighted in panel e by the dashed line. (f) The untwinning degree, determined  by the intensity of the measured Bragg peaks, is plotted as a function of the film thickness. 
} 
\end{figure*}

The next step to obtain underdoped films is the removal of oxygen from the CuO chains, so to get access to the underdoped side of the YBCO phase diagram. In the case of $c$-axis oriented YBCO films, we have succeeded by carefully tuning the in-situ post-annealing, i.e. cooling down the film soon after the deposition at a 5$^{\circ}/$min rate, with an oxygen pressure reduced from 900 mbar down to $1\times10^{-5}$ mbar, depending on the required doping level \cite{arpaia2018probing, andersson2020fabrication}. We have applied this procedure to the $a$-axis oriented films, and performed a transport characterization via resistance vs temperature $R$($T$) measurements. A typical result, for a 300 nm thick film in-situ annealed at a pressure of $1\times10^{-3}$ mbar, is shown in Figure \ref{fig:Fig2}(a). The zero resistance critical temperature is reduced down to 50 K, as expected for an underdoped film. However, the onset of the superconducting transition  is at $T \sim 91$ K, which hints to the presence of residual optimally doped domains. Therefore this annealing procedure, that we have successfully used for $c$-axis oriented YBCO films, cannot be implemented for $a$-axis oriented YBCO films, since it leads to films with inhomogeneous doping. The result is not surprising, considering the nanoribbon-like structure of the $a$-axis films: the oxygen planes are strongly exposed to the environment, which alter the oxygen diffusion conditions for achieving a homogeneous oxygen content across the whole film \cite{mori2011oxygen, baghdadi2014toward, baghdadi2015fabricating}. Here it is worth mentioning that performing a longer in-situ annealing at an intermediate temperature during the cooling does not improve the superconducting properties and the sample homogeneity. 
To overcome this issue, we have used an alternative approach, already successfully employed to modifying the oxygen doping in $c$-axis oriented films \cite{wuyts1993influence}: starting from a high quality slightly overdoped film, we decrease the oxygen doping with a separate step. The annealing, which changes the oxygen doping, is  performed ex-situ, in a reduced oxygen pressure (in the range of $10^{-5} - 10^{2}$ mbar, depending on the desired doping level) for 9 h at 650$^{\circ}$C. Keeping a constant pressure, the film is subsequently cooled down to room temperature using a constant cooling rate of 5$^{\circ}/$min. Figure \ref{fig:Fig2}(b) shows the results of this procedure: the film, annealed ex-situ using the same oxygen pressure as the film showed in Fig. \ref{fig:Fig2}(a), presents instead a rather sharp superconducting transition, reflecting the high level of homogeneity of the sample.

This latter procedure offers the advantage to be able to modify the doping of the same film several times. The whole underdoped side of the phase diagram can therefore be explored  using a single sample, in which only the oxygen content is modified.

\section{Structural characterization: complete detwinning and strain-induced orthorhombicity}

A thorough X-ray Diffraction (XRD) analysis has been performed {\color{black} using a four-circle Panalytical X’pert diffractometer with CuK$_\alpha$ radiation ($\lambda = 1.5406$ \AA), incorporating a hybrid Ge(220) monochromator and a PIXcel 3D detector matrix. The primary objective of the analysis is} to determine the structural properties of the ex-situ-annealed underdoped $a$-axis films.

The symmetrical $2\theta-\omega$ scans (see Fig. \ref{fig:Fig3}(a)) confirm what has been previously observed on the as-grown, fully oxygenated films, i.e. that the films are highly crystalline and $a$-axis oriented, with a negligible contribution coming from possible $c$-axis oriented grains. The homogeneity of the films is doping independent, as proved by the full width at half maximum FWHM of the (300) YBCO Bragg peak, $\Delta2\theta^{(300)}$, which remains constant from the slightly overdoped down to the undoped region (see left inset of Fig. \ref{fig:Fig3}(a)).

To get more insight into the twinning state of these systems, we have performed  symmetrical $2\theta-\omega$ maps in the $2\theta$ region around the (300) peaks of PBCO and YBCO (see Fig. \ref{fig:Fig3}(a)).
A third peak is visible, at intermediate angles between these two, whose intensity grows when the film thickness increases. This is the (030) YBCO reflection, associated to the $b$ axis: a small fraction of the $a$ and $b$ axis is randomly exchanged, i.e. a small $a$-$b$ twinning occurs. We get confirmation of that, performing asymmetrical  $2\theta-\omega$ maps around the (038)-(308) YBCO reflections along the $c$-axis direction (see Fig. \ref{fig:Fig3}(c)), and ascertaining the presence of a rather weak (038) peak. Rotating the sample by 90 degrees, and performing the same measurement along the $b$-axis direction, an extremely weak (308) reflection appears (see Figs. \ref{fig:Fig3}(d)-(e)). In this configuration, the detection of any ($\ell0\enn$) Bragg peaks should be forbidden, since the scattering plane has no component along the $c$-axis direction. This unexpected peak is therefore related to a random exchange, a twinning, of the $b$ and $c$ axis. From the intensity of the YBCO reflections in the symmetrical and asymmetrical maps, we have calculated the detwinning degrees as a function of the film thickness (see Fig. \ref{fig:Fig3}(f)).  The detwinning $b$/$c$ is thickness independent, and close to 99\%. The detwinning  $b$/$a$ gets instead smaller by increasing the film thickness from 50 to 800 nm, going from $\sim$97\% to $\sim$88\%. In particular for the thinnest films, this corresponds to a nearly complete detwinning, comparable to  that in single crystals, and better than any result achieved so far for YBCO thin films \cite{brotz2000detwinning, dekkers2003monocrystalline, arpaia2019untwinned}. 

\begin{figure}[thb!]
\centering
\includegraphics[width=8.5cm]{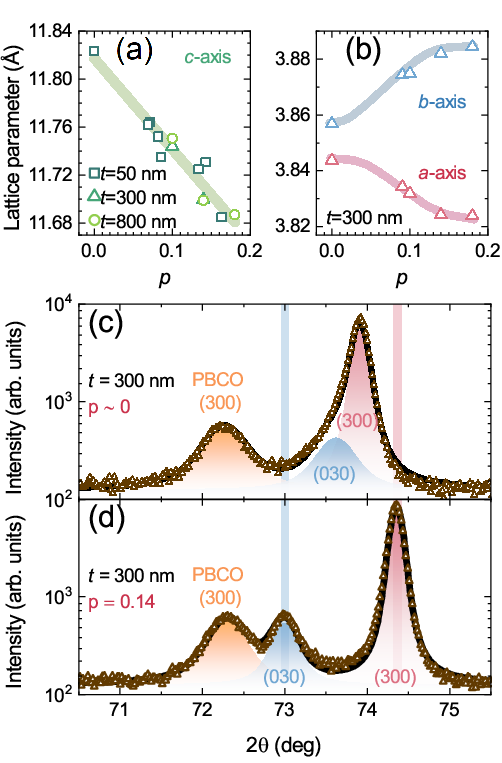}
\caption{\label{fig:Fig4}  Locking of the CuO$_2$ planes in $a-$axis oriented YBCO films. (a) The $c$-axis parameter, extracted from the ($\ell00$) and ($\ell0\enn$) Bragg reflections, is shown as a function of the doping $p$ for films of different thickness. {\color{black} For each sample, $p$ has been determined based on the $c$-axis length, and then utilizing the critical temperature $T_{\mathrm{c}}$ measured via transport (see Sec. \ref{sec:transport}) and the empirical parabolic relationship connecting doping and $T_{\mathrm{c}}$ in cuprate HTS \cite{liang2006evaluation, arpaia2018probing}.} Notably, all data points for the 50 nm thickness have been collected from a single, multiply-annealed thin film.  (b) The $a$- and $b$-axis parameters, extracted from the ($\ell00$) and ($\ell\emm0$) Bragg reflections, are plotted vs $p$ for different 300 nm thick films. (c)-(d) The symmetrical $2\theta-\omega$ scans respectively of an insulating ($p \sim 0$) and of an underdoped ($p = 0.14$) film are compared. The decomposition of the data into three Voigt profiles associated to the (300) PBCO, the (030) YBCO and the (300) YBCO reflections shows first that the PBCO layer is unaffected by the annealing procedure, as the peak position is unchanged vs doping. Secondly, that the two YBCO peaks, related to the CuO$_2$ planes, are distinct, even though they get closer, in the undoped sample. To facilitate the comparison, the vertical bars show the position of the two YBCO peaks in the doped sample.
} 
\end{figure}

From the positions of the YBCO reflections in the symmetrical and asymmetrical maps, we have extracted information about the structure of the thin films and on the modification of the unit cell induced by the strain and by confinement effects. The length of the $c$-axis parameter increases as the ex-situ oxygen pressure is decreased. This proves that our process is effective in varying the oxygen concentration within the films. The doping $p$ can therefore be estimated, by the combined knowledge of the $c$-axis parameter and of the $T_{\mathrm{c}}$, using a procedure already successfully used for YBCO single crystals \cite{liang2006evaluation} and thin films \cite{arpaia2018probing}. Finally, we can plot the $a$-, $b$- and $c$-axis lengths vs $p$ (see Figs. \ref{fig:Fig4}(a)-(b)). 
The YBCO $c$-axis is relaxed, with a length almost coinciding with the bulk value \cite{jorgensen1990structural} already at a thickness of 50 nm. 
When reducing the oxygen doping, a remarkable effect of the strain on the YBCO $a$- and $b$-axis parameter occurs. The orthorhombic to tetragonal transition, usually happening below $p \sim 0.06$ \cite{jorgensen1990structural}, is indeed absent. This is evident already from the symmetrical  $2\theta-\omega$ scans (see Figs. \ref{fig:Fig4}(c)-(d)): comparing an insulating to a slightly underdoped sample, one can notice that the (030) and (300) peak positions get much closer in the undoped case, but are still distinct. 

\begin{figure}[thb!]
\centering
\includegraphics[width=8.5cm]{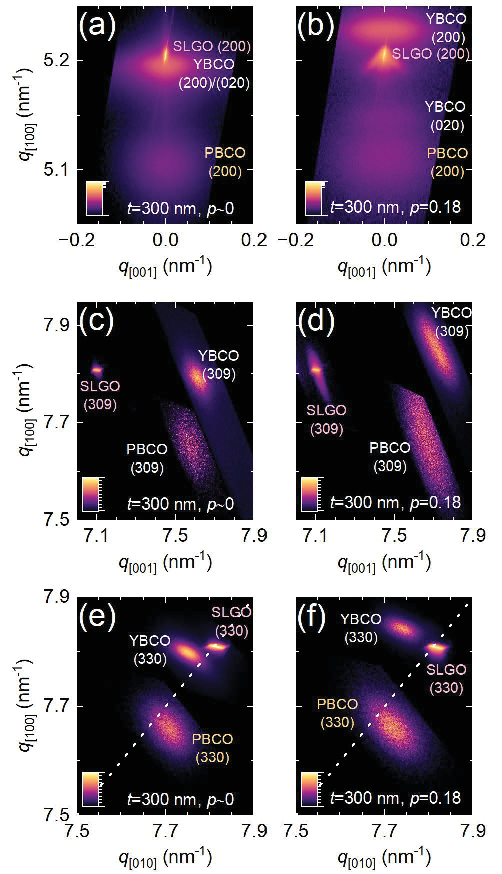}
\caption{\label{fig:RSM}  {\color{black} Suppression of the orthorhombic to tetragonal transition. (a)-(b) Symmetric and (c)-(d)-(e)-(f) Asymmetric Reciprocal Space Maps (RSM) of two 300 nm thick $a$-axis oriented YBCO films. Panels (a)-(c)-(e) refer to an undoped sample, while panels (b)-(d)-(f) refers to a fully oxygenated sample. (a)-(b) Symmetric map of the (200) reflections of the SLGO, PBCO and YBCO. (c)-(d) Asymmetric map of the (309) reflections of the SLGO, PBCO and YBCO. (e)-(f) Asymmetric map of the (330) reflections of the SLGO, PBCO and YBCO. In panels (e)-(f), the dashed line represents the tetragonal line $a=b$. The $a$, $b$ and $c$ axis parameters can be determined from the RSM peaks via  $a = \ell/q_{\mathrm{[100]}}$, $b = \emm/q_{\mathrm{[010]}}$ and $c = \enn/q_{\mathrm{[001]}}$, where $\ell$, $\emm$ and $\enn$ are the miller indices of the peak ($\ell\emm\enn$).}
} 
\end{figure}

{\color{black} 
To comprehensively address the strain mechanism in our films, at the origin of the suppression of the orthorhombic to tetragonal transition, we have examined two extreme doping cases, namely, near-zero doping and a doping level of 0.18. For each of these two samples, we conducted reciprocal space maps (RSM) of the (200), (309), and (330) reflections of SLGO, PBCO, and YBCO (see Fig. \ref{fig:RSM}), which also provided insights into the lattice parameters of the different layers (see Table \ref{RSMpar}).

The symmetric RSM in Figures \ref{fig:RSM}(a)-(b) show a perfect alignment of the YBCO (200) and PBCO (200) reflections with the SLGO (200) reflection at both doping levels, which supports the high texture of the films.

With knowledge of the out-of-plane $a$-axis parameter for both PBCO and YBCO, we determined the in-plane parameters and strain conditions by performing RSM of the asymmetrical (309) (see Figs. \ref{fig:RSM}(c)-(d)) and (330) (see Figs. \ref{fig:RSM}(e)-(f)) reflections. The maps confirm the pivotal role of the PBCO layer in determining the final crystal structure of the YBCO film. Firstly, the PBCO $c$-axis parameter mediates the tension between the substrate and the YBCO layer, accommodating the significant strain along the in-plane [001] direction (with a lattice mismatch of about 7-8\% between substrate and film). As we can see in Figures \ref{fig:RSM}(c)-(d), the PBCO layer exhibits a tensile strain along this direction, resulting in a broad (309) reflection, indicative of a mosaic structure, with the $c$-axis parameter expanding by 0.5\% relative to bulk values. This allows the relaxation of the YBCO $c$-axis toward bulk parameters at any doping levels (as also observed in Fig. \ref{fig:Fig4}(a)), reducing the mosaicity along that direction. Secondly, the PBCO $b$-axis parameter significantly influences the final structure of the YBCO layer. This occurs because its length is close, though slightly longer, than those of SLGO and YBCO, and weakly doping-dependent compared to the YBCO $b$-axis. Independently of the doping level, the PBCO $b$-axis is compressively strained by the substrate to a value ($b = 3.89$ Å) that is very close to that of slightly overdoped YBCO. Consequently, a doping-dependent strain is induced by the PBCO into the YBCO layer. In the $p=0.18$ sample (see Fig. \ref{fig:RSM}(f)), the PBCO and YBCO (330) reflections have almost identical $q_{\mathrm{[010]}}$ component: the YBCO $b$-axis  is subject to a tiny strain and preserves its bulk value. In the undoped sample (see Fig. \ref{fig:RSM}(e)), the tensile strain applied by the PBCO layer maintains the YBCO $b$-axis longer than the $a$-axis (which undergoes shrinkage due to the Poisson effect, see Table \ref{RSMpar}), leading to the disappearance of the orthorhombic-to-tetragonal transition. This evidence is clearly depicted in Figure \ref{fig:RSM}(e), where the YBCO (330) reflection remains notably distant from the dashed line corresponding to the tetragonal line $a=b$.  
}

\begin{table*}[hbt!] 
{\color{black}
\begin{tabular}{c c c c c} 
\hline \hline
 \; \;\; Material \;\; \;  &   \;\; \; $a$ (Å) \; \; \; &  \;\; \; $b$ (Å)  \;\; \; &  \;\; \;  $c$ (Å) \;\; \; &  \;\; \;  $V$ (Å$^3$) \;\; \; \\
\hline
SrLaGaO$_4$ & 3.842 & 3.842 & 12.678 & 187.140 \\
PrBa$_2$Cu$_3$O$_6$ & \; 3.915 (3.906) \; & \; 3.890 (3.906) \; & \; 11.88 (11.824) \; & \; 180.92 (180.40) \; \\
PrBa$_2$Cu$_3$O$_7$ & \; 3.916 (3.913) \; & \; 3.885 (3.916) \; & \; 11.76 (11.712) \; & \; 178.91 (179.46) \; \\
YBa$_2$Cu$_3$O$_6$ & \; 3.848 (3.86) \; & \; 3.869 (3.86) \; & \; 11.82 (11.817) \; & \; 175.98 (176.07) \; \\
YBa$_2$Cu$_3$O$_7$ & \; 3.825 (3.819) \; & \; 3.885 (3.886) \; & \; 11.68 (11.68) \; & \; 173.57 (173.34) \; \\
\hline
\end{tabular}
\caption{Summary of the lattice parameters, related to the 300 nm thick YBCO films, extracted from the maps in Figure \ref{fig:RSM}. In parenthesis, the bulk lattice parameters of YBCO (from \cite{jorgensen1990structural}) and PBCO (from \cite{lowe1992search} and \cite{uma1996structural} respectively for the undoped and doped compound) are reported. In the last column, the volume $abc$ of the various unit cells is reported,  confirming the Poisson effect in our samples.} \label{RSMpar}
}
\end{table*}

In conclusion, the strain has important consequences on the crystal structure: the CuO$_2$ planes, that are confined by the nanoscale film thickness along $a$, are also inflenced by the substrate {\color{black} and seed layer}, which lock the value of the $b$-axis.  What is remarkable is that this locking also occurs in films, {\color{black} with thicknesses significantly exceeding the relaxation threshold observed in $c$-axis oriented YBCO films \cite{contour1996critical} (see Figs. \ref{fig:Fig4}(b) and \ref{fig:RSM}(e)).}

\section{Transport characterization} \label{sec:transport}

To measure the resistance vs temperature $R$($T$) along the two in-plane directions of the films, we employed very elongated Hall bar-like structures, realized by a soft patterning procedure we have developed over years for various cuprate HTS \cite{baghdadi2017study, andersson2020fabrication, charpentier2016hot}. The current is applied at the two extremities of the bars, with the two voltage probes positioned 10 squares apart.
\begin{figure}[htb]
\centering
\includegraphics[width=8.5cm]{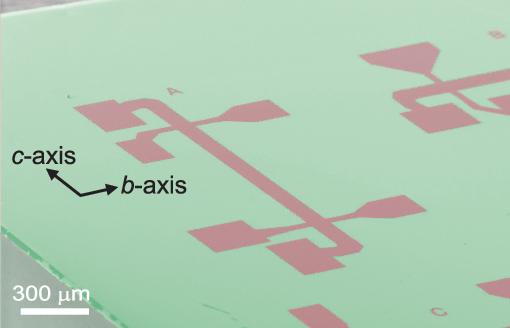}
\caption{\label{fig:FigSEM}  Scanning Electron Microscopy image of the structures used for the measurements of the nanoribbon resistivity. The 1 mm long and 100 $\mu$m wide Hall bars are oriented along either the $b$-axis or the $c$-axis direction of the YBCO film.
} 
\end{figure}
The choice of this geometry is motivated by  the high in-plane resistivity anisotropy expected in our films, due to the weak coupling between adjacent CuO$_2$ planes, i.e., along the $c$-axis direction. Consequently, the standard  Van der Pauw method \cite{philips1958method}, which assumes isotropic resistivity and homogeneous resistivity distribution, cannot be applied. 

\begin{figure}[htb]
\centering
\includegraphics[width=8.5cm]{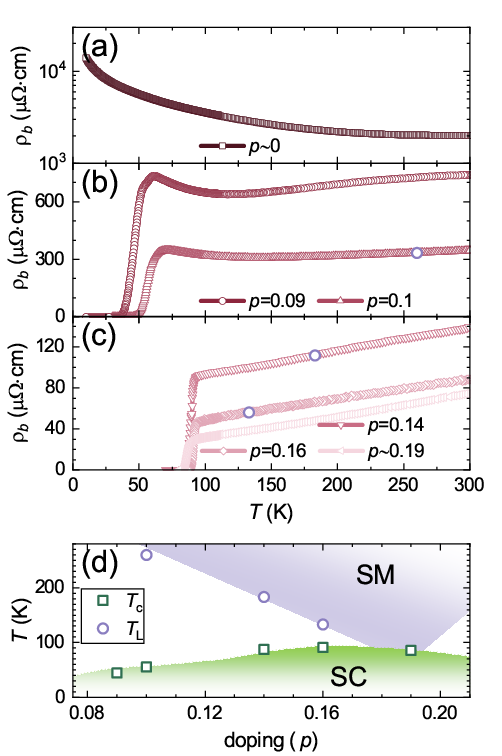}
\caption{\label{fig:Fig5}  Doping dependence of nanoribbon resistivity. Panels (a)-(b)-(c) show $\rho_b$($T$) for various  300 nm thick films at different oxygen doping levels. The critical temperature $T_{\mathrm{c}}$ for each film is determined from the maximum of the first derivative of $\rho_b$($T$). $T_{\mathrm{L}}$ (violet circles), marking the crossover between the pseudogap region and the strange metal phase, {\color{black}is the temperature below which the resistivity $\rho_b$($T$) deviates by 1\% from the high temperature linear fit of the curves, performed in the interval 280 K$<T<$300 K. Notably, all $\rho_b$($T$) datasets are extensive, featuring a sub-kelvin temperature step-size.} Panel (d) presents a phase diagram constructed from the obtained values of $T_{\mathrm{c}}$ and $T_{\mathrm{L}}$, exhibiting good agreement with the superconducting (SC) and strange metal (SM) regions observed in YBCO single crystals and $c$-axis oriented films \cite{barivsic2013universal, arpaia2018probing}.   
} 
\end{figure}

We have performed our measurements on samples with thicknesses in the range between 50 and 800 nm, and with doping levels going from the undoped up to the slightly overdoped. In the following, we will focus however only on the results related to the 300 nm thick films. An overview of the resistivity measurements vs temperature along the $b$-axis direction, $R_b$($T$), are shown in Figures \ref{fig:Fig5}(a)-(b)-(c).
The curves refer to different doping levels and are characterized by rather sharp superconducting transitions and by a normal state resistance, qualitatively following the typical  dependencies  observed in  $c$-axis oriented films (see Fig. \ref{fig:Fig5}(d)) \cite{barivsic2013universal, arpaia2018probing, arpaia2017transport, wuyts1993influence, sefrioui1999crossover}. This refers in particular to the doping dependent crossover temperature $T_{\mathrm{L}}$, inferred by the departure from the linear-in-$T$ behavior of the resistance at high temperature, which appears as a sort of boundary between the unconventional strange metal phase and a region where the material loses low energy electronic excitations (the pseudogap region). {\color{black} A summary of the parameters extracted for the films of Figure \ref{fig:Fig5} is shown in Table \ref{UDTable}.} 

\begin{table}[hbt!] 
{\color{black}
\begin{tabular}{c c c c} 
\hline \hline
 \; \; $p_{\mathrm{ann}}$ (mbar) \; \;  &   \; \; $T_{\mathrm{c}}$ (K)  \; \; &  \; \; $T_{\mathrm{L}}$ (K)  \; \; &  \; \;  $p$ \; \; \\
\hline
$6.0 \cdot 10^{-5}$ & - & - ($>$ 280) & 0 \\
$3.2 \cdot 10^{-4}$ & 45 & - ($>$ 280) & 0.09 \\
$1.0 \cdot 10^{-3}$ & 56 &  260 & 0.10 \\
$1.1 \cdot 10^{-2}$ & 88 & 183 & 0.14 \\
$1.0 \cdot 10^{0}$ & 91 & 133 & 0.16 \\
$9.0 \cdot 10^{2}$ & 86 & - & 0.19 \\
\hline
\end{tabular}
\caption{Summary of the parameters related to the 300 nm thick films presented in Figure \ref{fig:Fig5}. The films have been achieved by varying the ex-situ annealing pressure $p_{\mathrm{ann}}$ across a seven orders of magnitude range, spanning from fully oxygenated to undoped samples. From the $R(T)$ of the films at different doping levels we have extracted $T_{\mathrm{c}}$ and $T_{\mathrm{L}}$, following the procedure described in the caption of Figure \ref{fig:Fig5}. Finally, for each film the doping level $p$ has been determined by the combined knowledge of $T_{\mathrm{c}}$ and of the $c$-axis length \cite{liang2006evaluation, arpaia2018probing, arpaia2023signature}.} \label{UDTable}
}
\end{table}

\begin{figure}[htb]
\centering
\includegraphics[width=8.5cm]{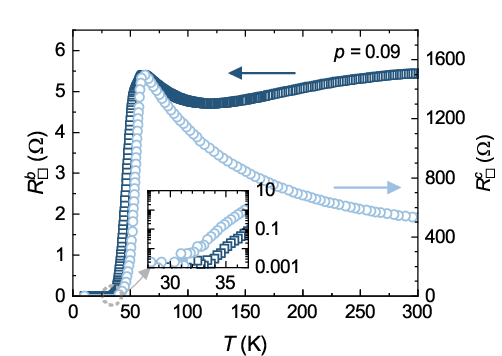}
\caption{\label{fig:Fig7}  Anisotropic transport properties of underdoped 300 nm thick CuO$_2$ nanoribbons. Sheet resistances $R_{\Box}^{b} $($T$) (squares) and $R_{\Box}^{c} $($T$)  (circles) of an underdoped ($p = 0.09$) film. The inset, where the tails of the superconducting transition of the two measurements are presented on the same vertical scale, highlights the lower zero resistance critical temperature occurring along the $c$-axis.
} 
\end{figure}

Figure \ref{fig:Fig7} shows instead the $R$($T$), along both the in-plane directions, of a typical 300 nm thick film, at a doping level ($p \sim 0.09$) where the CuO$_2$ planes - here confined at the nanoscale - have been shown to host both spin \cite{haug2010neutron} and charge density waves \cite{ghiringhelli2012long}. 
In the normal state, a resistivity anisotropy much stronger than in optimally doped YBCO ($R_c/R_b \approx 60$ at room temperature \cite{friedmann1990direct, takenaka1994interplane}) is present. This is expected, as a consequence of the weaker coupling between the CuO$_2$ planes, occurring in the underdoped regime because of the increased $c$-axis length, i.e. of the increased distance between the planes. {\color{black} In our case, at room temperature, $R_c/R_b \approx 70$ for the optimally doped films; when decreasing the oxygen doping,}  $R_c$ becomes $\sim$100 times larger than $R_b$. This anisotropy, comparable to the best results on detwinned single crystals \cite{takenaka1994interplane, zverev2000anisotropy}, is reported for the first time in {\color{black} underdoped} $a$-axis oriented YBCO thin films.  
This suggests that the coupling between adjacent CuO$_2$ planes is weak already in rather thick films, as also highlighted by the insulator-like temperature dependence of $R_c$, and that it can be possibly made even weaker in few-unit-cell thick films, where the effect of the confinement is enhanced. 

Furthermore,  as shown in the inset of Fig. \ref{fig:Fig7}, the zero resistance critical  temperature is also anisotropic, being lower along the $c$-axis direction. This suggests a weaker interlayer Josephson coupling along the $c$-axis, reminiscent of the observations  in underdoped `214' cuprates like La$_{2-x}$Ba$_x$CuO$_4$ \cite{tranquada2008evidence}. In this rather special cuprate family, this occurrence  has been theoretically associated to the formation of stripes, where CDW and SDW are coupled, and to the presence of pair density waves (PDW) \cite{agterberg2020physics}. The appearance of a similar phenomenology in  YBCO might indicate the potential influence of confinement on the material ground state, with modified CDW and SDW compared to the bulk. The investigation of these local orders in underdoped CuO$_2$ nanoribbons will be object of future work, extending beyond the scope of the present study.

\section{Conclusions}
We have presented the fabrication and characterization of underdoped CuO$_2$ nanoribbons, realized by employing $a$-axis oriented YBCO thin films. The oxygen doping is tuned by using an ex-situ annealing procedure, which allows the achievement of films covering a broad region of the HTS phase diagram, going from the undoped insulator up to the slightly overdoped level. Taking advantage of the fast oxygen diffusion in the $a-b$ planes, which are open in our films, this procedure demonstrates its effectiveness to vary the oxygen doping up to high thicknesses (of the order of 800 nm). Structural XRD characterization demonstrates that the films are nearly totally detwinned, and the strain affects the crystal structure, up to the largest investigated thickness: the CuO$_2$ planes have a nanoribbon shape, which inhibits the orthorhombic-to-tetragonal transition and possibly induces the system to be more anisotropic than in the bulk. The homogeneity of the films has been demonstrated by $R$($T$) measurements, showing rather narrow superconducting transitions. The transport properties exhibits notable in-plane anisotropy in both  normal state resistance and  critical temperature,  indicative of  weak coupling between the CuO$_2$ planes.

The system we have engineered, where CuO$_2$ planes are confined into nanoribbons, stands out as a natural candidate for studying local orders in HTS at the nanoscale. In this setting, the effects of strain and confinement may simplify the underlying physics, offering valuable insights into the intricacies of these materials.

More in general, this platform holds significant promise for advancing the understanding of HTS in both fundamental physics and practical applications. 
In terms of basic physics, $a$-axis oriented films are inherently suitable for pump probe experiments. The intense electric field of mid-infrared light pulses can align with the crystallographic $c$-axis, which overcomes limitations in $c$-axis oriented samples, where electric field interactions are restricted to in-plane excitations. The electronic structure of cuprates and the position of apical oxygen atoms get  therefore strongly altered  \cite{disa2021engineering}, which can provide valuable insights into the phenomenology behind light-induced superconductivity  \cite{buzzi2018probing}.
On the practical side,  films oriented along the $a$-axis, where the superconducting coherence length is approximately an order of magnitude longer than along the $c$-axis, offer an ideal system for fabricating high-performance Josephson junctions, which are crucial components for superconducting electronics applications \cite{makhlin2001quantum, blais2021circuit}.

\section*{Acknowledgements}
The authors acknowledge Andrea Cavalleri and Michael Först for insightful discussions, Henrik Frederiksen for technical support. This work was performed in part at Myfab Chalmers, and is supported by the Area of Advance Nano program at Chalmers, the 2D Tech VINNOVA competence Center (2019-00068) and the Swedish Research Council (VR) under the projects  2020-04945 (R.A.), 2020-05184 (T.B.) and 2022-04334 (F.L.).

\bibliography{bibl}

\end{document}